\documentclass[prb,aps,twocolumn,showpacs,floatfix]{revtex4}
\usepackage{graphicx}

\begin{document}
\title{Interplay between SDW and induced local moments 
in URu$_2$Si$_2$
} 
\author{V. P. Mineev and M. E. Zhitomirsky}
\affiliation{Commissariat \'a l'Energie Atomique, DSM/DRFMC/SPSMS 
38054 Grenoble, France}
\date{December 9, 2004}

\begin{abstract}
Theoretical model for magnetic ordering in the heavy-fermion metal 
URu$_2$Si$_2$ is suggested. The 17.5~K transition in this material
is ascribed to formation of a spin-density wave, 
which develops due
to a partial nesting between electron and hole parts of the Fermi
surface and has a negligibly small form-factor. 
Staggered field in the SDW state induces
tiny antiferromagnetic order in 
the subsystem of 
localized singlet-singlet levels. Unlike the other models our scenario 
is based on coexistence of two orderings with the same
antiferromagnetic dipole symmetry.
The topology of the pressure phase diagram
for such a two order parameter model is studied in the
framework of the Landau theory.
The field dependences of the staggered magnetization and 
the magnon gap are derived from the microscopic theory and  
found to be in good quantitative agreement with experiment.
\end{abstract}
\pacs{
71.27.+a,  
75.10.-b,  
75.30.Fv   
}
\maketitle

\section{Introduction}

URu$_2$Si$_2$ is one of the most intriguing heavy-fermion compounds. 
It exhibits a sharp second-order transition at $T_m=17.5$~K, which has
pronounced effect on thermodynamic and kinetic properties,
\cite{palstra85,schlabitz,maple86,palstra86,visser} though the 
neutron diffraction experiments \cite{broholm87,mason90,broholm91,fak96} 
and the $X$-ray magnetic scattering 
measurements \cite{isaacs} have produced evidence for
only tiny staggered moments $\mu\simeq 0.02$--$0.04\mu_B$ 
at ${\bf Q}=(1,0,0)$.
Magnetic excitations observed by inelastic neutron 
scattering experiments \cite{broholm87,broholm91} are 
reasonably well explained within the model with exchange interaction in a
singlet-singlet Van-Vleck paramagnet.\cite{nieuwenhuys} 
This model fails, however, to give a consistent description 
of the ordering temperature and small ordered moments 
unless the exchange interaction ${\cal J}_{\bf Q}$ is accidentally close to a 
critical value (see section IV below).
A weak antiferromagnetic ordering of a Van Vleck paramagnet
cannot produce a measured jump in the specific heat \cite{maple86} 
and an electrical resistivity anomaly. \cite{palstra86}  
These experimental features rather resemble formation of a
spin-density wave (SDW), which involves approximately 
a half of the Fermi surface.
In its turn, the SDW scenario is inconsistent with a longitudinal
polarization of the magnetic excitations. 
As a result of this contradiction, various theoretical scenarios
have been proposed in the past to explain intriguing behavior of
URu$_2$Si$_2$.  \cite{gorkov,ramirez,barzykin,agterberg,santini,sikkema,
okuno,ikeda,shah,chandra,fazekas,fomin}

The proposed theories of URu$_2$Si$_2$ can be crudely divided 
into three broad categories.
Phenomenological models of the first group \cite{agterberg,shah}
take pragmatic approach and 
introduce a yet unknown primary or hidden order parameter $\psi$,
which drives the 17.5~K transition and is responsible for 
a large specific heat anomaly at $T_m$.
Small ordered moments observed in the neutron experiments are described by 
a secondary Ising-like antiferromagnetic order parameter $m$.
Depending on the symmetry of the hidden order parameter
different coupling terms of the type $\psi m$ and $\psi^2m^2$ are possible
in the Landau energy functional.
In particular, the model with a  bilinear term predicts
a nontrivial field-dependence of weak antiferromagnetic moments with an inflection
point. \cite{shah}  
Subsequent neutron measurements have nicely  confirmed such a prediction.\cite{bourdarot03}

Second group of theories consists of specific proposals for the hidden order 
parameter. These models are further subdivided into two subgroups.
The first subgroup includes models, in which the primary order parameter 
breaks the time-reversal symmetry. These are spin-density waves 
in higher angular momentum channels, \cite{ramirez,ikeda} triple 
spin correlators,\cite{gorkov,agterberg} orbital antiferromagnetism, 
\cite{chandra} and ordering of octupolar moments on uranium 
sites. \cite{fazekas} 
For such theories the bilinear coupling term 
with the antiferromagnetic vector is, in principle, possible, though 
some other crystal symmetries are required to be absent.
The second subgroup includes theoretical models with a time-reversal
invariant hidden order parameter such as quadrupolar \cite{santini}
or spin  nematic ordering.\cite{barzykin} 
In such a case the bilinear term is prohibited by symmetry
and only a biquadratic interaction exists between the two
order parameters.

The last third group of theoretical works includes attempts to 
realistically describe the microscopic interactions in
URu$_2$Si$_2$.
A so called Ising-Kondo lattice model \cite{sikkema}  
describes interaction between conduction electrons in 
nested bands and local 
crystal-field split moments.
The mean-field calculation 
produces both a weak moment and an appropriate value of the transition 
temperature but does not reproduce the large specific heat jump. 
A closely related dual model \cite{okuno} 
considers a subsystem of localized singlet-singlet levels and
a subsystem of itinerant electrons with a similar assumption on the 
nesting condition. 
A better description of the thermodynamic
properties of URu$_2$Si$_2$ has been achieved by adding the superexchange
and the RKKY interactions between local
moments. The field behavior still remains largely 
inconsistent with the experimental measurements.\cite{bourdarot03} 
Further development of the dual model has been recently 
suggested in Ref.~\onlinecite{fomin}.
Note, that the above two microscopic models (i) completely neglect 
the Coulomb interaction between charge carries and (ii) operate with a single
(antiferromagnetic) order parameter.

Up to now experiments have been unable to distinguish
between the competing theoretical proposals.
Two potentially important experimental developments 
published recently include pressure experiments and the NMR measurements.
Investigations under hydrostatic pressure\cite{amitsuka,motoyama,bourdarot}
have shown that 
the $P$--$T$ phase diagram is divided into two regions: 
a small moment antiferromagnetic phase (SMAF) at low pressures 
and a large moment antiferromagnetic phase (LMAF) at high 
pressures with a first-order transition line in between.  Such 
a discovery is generally consistent  with phenomenological two order
parameter scenarios for URu$_2$Si$_2$ (see section III, below).
Another experimental finding came out from the NMR measurements.
Matsuda and co-workers\cite{matsuda01,matsuda03} have found
that a paramagnetic Si$^{29}$ NMR absorption line persists well 
below $T_m$ accompanied by two much smaller peaks symmetrically 
shifted by antiferromagnetic field. Such an observation points at
an inhomogeneous para-antiferromagnetic state below the 17.5~K transition.  
The peak intensities suggest that about 97\% of the sample 
volume is in a paramagnetic state.  
The puzzle of small uranium moments in URu$_2$Si$_2$ seems to be
reinterpreted as due to a phase separation
between nonmagnetic state with a hidden order and small 
antiferromagnetic droplets with ordinary (large)
value of staggered magnetization. Such a simple explanation of 
the main mystery of URu$_2$Si$_2$ is quite appealing  
but leaves without answer question why
the inhomogeneous phase exists not only in the vicinity of the first-order
transition line at high pressures,
but in the whole region of SMAF. 
The development of the antiferromagnetic Bragg peaks right below 
$T_m=17.5$~K seems to be highly accidental in the phase separation
scenario. Also, an inflection point in the field
dependence of antiferromagnetic Bragg peaks cannot be explained 
if small peaks are purely due to a volume effect. Thus, physical 
implications of the NMR measurements \cite{matsuda01,matsuda03} 
are not completely straightforward and have to be further clarified.

In this paper, we present a semimicroscopic model for 
URu$_2$Si$_2$, which 
is closely related to the above mentioned dual
models,\cite{sikkema,okuno} but operates with two order parameters
in the spirit of phenomenological scenarios.\cite{agterberg,shah}
We also consider two magnetic subsystems: (i) local crystal-field
split moments 
on U$^{4+}$ sites and (ii) conduction electrons in nested 
bands. In contrast to the previous works\cite{sikkema,okuno}
we assume that the electron-electron interaction is nonnegligible
and that it drives a SDW transition in the nested parts of the Fermi surface. 
The critical temperature
$T_m=17.5$~K is associated with $T_{\rm SDW}$ and the SDW
amplitude $\psi$ plays the role of a primary (hidden) order parameter. 
According to the LSDA calculations \cite{norman} the nesting
wave-vector is commensurate and
corresponds to the experimentally observed two-sublattice
antiferromagnetic structure. The SDW formed in conduction 
bands is responsible for large changes in thermodynamic 
and kinetic properties of URu$_2$Si$_2$. At the same 
time, we argue that the SDW has a small form-factor and does not create
significant Bragg reflection.  Local polarization of uranium sites by
a staggered magnetic field from the SDW induces tiny antiferromagnetic
moments.  The magnetic dynamics probed by neutrons is also determined
by a localized subsystem.

By making a single assumption about 
a hidden order parameter with the same symmetry as
the observed antiferromagnetic ordering,
we have further derived several results, which allow detailed comparison with
available measurements and suggest future experimental tests:  

(i) The $P$--$T$ phase diagram with line of the first
order type transition, which terminates at the critical point  below
$T_m(P)$; 

(ii) Field dependence of staggered
magnetization coincides perfectly with the experimental data
for a realistic set of microscopic parameters; 

(iii) Field dependence of the excitation spectrum. 

The  article is organized as follows.  In section II, 
we introduce the model and discuss various
features of the spin-density wave instability specific for
URu$_2$Si$_2$. In section III, using the phenomenological
approach with an appropriate Landau energy functional for
two coexistent order parameters
we investigate the topology of the $P$--$T$ phase diagram
of URu$_2$Si$_2$. 
In the next section, a system of localized crystal-field split
singlet-singlet levels is considered under combined influence of a
uniform external magnetic field and a staggered ``internal'' field
induced by the SDW. We calculate the field dependences of the
staggered magnetization and the excitation spectrum.  Comparison to the
experimental data is presented in section V, which is followed by
discussion and conclusions.

\section{Spin-density wave}

An unusual assumption made by phenomenological 
theories \cite{agterberg,shah} is a description of URu$_2$Si$_2$ 
with two order parameters having the same
symmetry. On the first site such an assumption contradicts 
to a general spirit of the Landau theory of phase transition.
Recent investigations of a two-gap superconductor MgB$_2$  
have, however, demonstrated usefulness of the description of the
superconducting state by means of two weakly interacting
$s$-wave condensates of the Cooper pairs  
(see, for instance, Ref.~\onlinecite{mzh}).  
The necessary condition for this is a significant mismatch of
the pairing interactions in the two bands. In the absence of 
interband scattering of the Cooper pairs, each band has its own superconducting 
transition temperature. An interband interaction is always present 
in real metals and leads to a single transition into a state
with two different gaps. The two gaps (order parameters)
still keep different dependences on temperature, 
pressure and/or applied magnetic field.
In relation to URu$_2$Si$_2$, the two order parameters
$\psi$ and $m$ should correspond to two significantly different
magnetic subsystems.  We suggest here that 
the primary order parameter may be
an ordinary SDW. The common objections against a SDW transition 
are (i) smallness of ordered moments and (ii)
longitudinal polarization of sharp magnetic excitations.
This two properties
can be reconciled with a SDW scenario by taking
into account specific features of URu$_2$Si$_2$.
In this section we discuss the former feature, while
the properties of magnetic excitations are considered
in section IV.

The early experimental works on the specific heat 
\cite{maple86,fisher90} and the magnetoresistance \cite{palstra86}
in URu$_2$Si$_2$ have found strong evidences in favor
of a charge or a spin-density wave instabilities
in the heavy-electron subsystem at
$T_m=17.5$~K. The fit of the electronic specific heat
below the transition indicates that 
a gap $\Delta_0\simeq 130$~K develops on 
40\% of the Fermi surface at $T\rightarrow 0$. 
This conclusion has received a strong support from  
the de Haas van Alphen (dHvA) measurements.\cite{ohkuni} 
Comparison of the measured dHvA frequencies to the {\it ab-initio}
band structure shows that two large 
pieces of the Fermi surfaces,
band-18 hole and band-19 electron, are not observed at low
temperatures, probably due to their partial removal below 
the ordering temperature. The above two sheets 
have nearly spherical shapes and are separated by
a nesting wave-vector ${\bf Q} = (0,0,1)$, which is
equivalent to $(1,0,0)$ in the Brillouin zone of a body-centered
tetragonal lattice.\cite{ohkuni} 
Direct calculation of a static momentum-dependent
susceptibility\cite{norman} also shows a peak 
at the commensurate wave-vector ${\bf Q}=(1,0,0)$.

A fast decrease of the uniform susceptibility\cite{ramirez} below $T_m$ 
as well as suppression of the transition temperature $T_m$ and 
the bulk gap $\Delta_0$ by applied magnetic field 
\cite{mentink,vandijk,jaime} also point to a charge- or 
a spin-density wave state.
For the CDW the Zeeman splitting degrades the nesting of 
the Fermi surfaces and reduces a mean-field transition 
temperature\cite{mckenzie} in a way, which is analogous 
to the paramagnetic limit effect in superconductors.
By contrast, an isotropic SDW involves coupling of bands
with opposite spin and the nesting is not affected by a magnetic
field. A strong spin-orbit coupling in heavy-fermion
materials creates momentum dependence of the $g$-factor.
If the nesting condition 
$\varepsilon({\bf k}+{\bf Q}) =-\varepsilon({\bf k})$ is satisfied 
for particular sheets of the Fermi surface it is not generally
fulfilled for the Zeeman shift 
$\mu_B g({\bf k}+{\bf Q})H/2\ne \mu_B g({\bf k})H/2$. 
Hence, in metals with strong spin-orbit coupling
an SDW state is also suppressed by the paramagnetic effect.

The mean-field theory of a SDW formation in ideally nested 
electron and hole Fermi surfaces\cite{gruner} resembles to
a large extent the BCS theory.
The relative jump of the specific heat at $T_m=T_{\rm SDW}$ 
is estimated by the BCS value 
$\Delta C/C\simeq 1.43$ ,
which is
compatible with the experimental value $\Delta C/C \approx 2.9$
once additional strong-coupling effects are 
taken into account.
The modulation of the spin density at $T=0$  
are given by
\begin{equation}
M^z_{\bf Q} = \mu_B \sum_{\bf k} \langle c^\dagger_{{\bf k}+{\bf
Q}\uparrow} c^{_{}}_{{\bf k}\uparrow} \rangle = \mu_B N_0 \Delta_0
\ln \frac{\varepsilon_F}{\Delta_0} \ ,
\label{szQ}
\end{equation}
where $N_0$ is the density of states per one spin direction.
Estimating $N_0\simeq n_e/\varepsilon_F$, we find that ordered 
moments normalized per one electron constitute a small fraction 
of the Bohr magneton $\sim \Delta_0/\varepsilon_F$. \cite{gruner} 
Such a reduction has a transparent physical meaning:
only electrons (holes) within a thin layer of width $2\Delta_0$ 
around the Fermi surface participate in the formation of ordered moments.
Thus, affecting strongly thermodynamic and
kinetic properties, a weak-coupling SDW order has a  small
form-factor and does not produce significant magnetic Bragg
scattering.
This fact  has not been so far appreciated in
the literature on URu$_2$Si$_2$. 

There are several additional factors, which complicate
the simple picture drawn above. First, the perfect nesting
between different bands does not appear in real materials. 
Absence of perfect nesting acts as a depairing effect reducing
gradually both the transition temperature and the zero-$T$ gap
and enhancing the residual density of states.
Obviously, such an effect does not change the conclusion about a small
form-factor, but may reduce the jump in the specific heat compared
to the BCS value.  In order to show that partial nesting does not modify the
previous estimate, we refer to a similar situation in superconductors
with paramagnetic (depairing) impurities.  Using the Abrikosov-Gor'kov
theory, Skalski {\it et al\/}.\cite{skalski} have calculated the
effect of paramagnetic impurities on various characteristics of an
$s$-wave superconductor.  Their results indicate that in a wide range
of impurity concentration, the jump in the specific heat and the
transition temperature are suppressed at approximately the same rate,
hence, preserving the BCS estimate for the 
relative specific heat jump. Second, the
electron mass enhancement in heavy fermion materials 
($m^*/m\sim 25$ in URu$_2$Si$_2$) can significantly
reduce the Fermi energy scale $\varepsilon_F$.  However,
simultaneously with a mass renormalization, an interaction with spin
fluctuations reduces strongly the spectral weight of heavy
quasiparticles\cite{varma,norman87} adding an extra small factor
$(m/m^*)^3$ to Eq.~(\ref{szQ}), which completely compensates
the factor $(m^*/m)$ in the density of states and further
reduces value of the ordered moments. Finally, according to the 
band structure calculations \cite{norman,ohkuni} URu$_2$Si$_2$ is 
a compensated metal with equal number of electrons and holes.
The number of carriers in the two bands undergoing a SDW transition
is smaller than one, when normalized to the number of U-atoms.
This yields an extra reduction factor, since
the neutron scattering experiments report the 
ordered moments normalized per one uranium.

The above arguments can, in our view, convincingly explain why small 
antiferromagnetic Bragg peaks in URu$_2$Si$_2$ are consistent
with a SDW instability.  In the following we assume that due to
nesting between some parts of the Fermi surface in URu$_2$Si$_2$ the
commensurate SDW state is formed below 
the critical temperature $T_m$ and that the
SDW amplitude $\psi\sim \sum_{\bf k}\langle c^\dagger_{{\bf k}+{\bf
Q}\uparrow} c^{_{}}_{{\bf k}\uparrow}\rangle$ plays the role of a
hidden order parameter in the problem.

\section{Phase diagram under pressure}

\begin{figure}[t]
\begin{center}
\includegraphics[width=0.99\columnwidth]{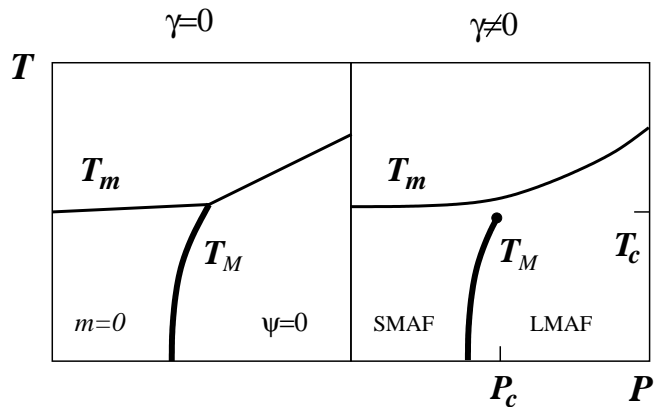}
\end{center}
\caption{
The phase diagram of the two-order parameter Landau functional
for $\gamma =0$ (left panel) and for 
$\gamma \neq 0$ (right panel).
}
\label{diagram}
\end{figure}

The microscopic dual models \cite{sikkema,okuno} assume that
conduction electrons are noninteracting and operate, therefore, with a
single order parameter, which leaves no place for the phase diagram
with SMAF and LMAF regions.  In our scenario, temperatures of
intrinsic phase transitions in itinerant and localized magnetic
subsystems are different functions of $P$ and they may interchange
their order under pressure.  As a result, a line of first-order
transitions appears naturally between the two ordered states, where
one order parameter prevails over another, see Fig.~\ref{diagram}.

The Landau free-energy for two interacting order
parameters can be written as
\begin{equation}
F=\alpha_1\psi^2 + \alpha_2m^2 + 2\gamma\psi m +
\beta_1\psi^4 + \beta_2m^4 + 2\beta_i \psi^2m^2 \,.
\label{landau}
\end{equation}
A special bilinear coupling term is allowed only if 
the two parameters transform according to the
same irreducible representation, otherwise $\gamma\equiv 0$.  For
nonzero $\gamma$, the quantities $\psi$ and $m$ do not correspond to
two different types of symmetry breaking.  Rather they describe two
weakly coupled magnetic subsystems of URu$_2$Si$_2$ in a way, which is
reminiscent of the Ginzburg-Landau description of the multigap
superconductivity in MgB$_2$.\cite{mzh}
The bilinear term corresponds, then,  
to a polarization of local moments by a spin-density wave.

Generally, in addition to the bilinear term $\psi m$ 
there are possible other coupling terms  in the 
Landau functional: $\psi^3m$ and $\psi m^3$.
These terms can exist even if $\psi$ and $m$ transform
according to different irreducible representations
(though $\psi$ has to break the time-reversal symmetry).
The $\psi^3 m$ term leads, for example, to a small 
antiferromagnetic component in a state with $\psi\neq 0$,
even if $\gamma\equiv 0$. 
The induced $m$ component grows in such a case
as $m \sim (T_m-T)^{3/2}$, while the neutron diffraction
experiments find a standard mean-field exponent 
$1/2$.\cite{broholm87,mason90}
This observation suggests that the $\psi m$ coupling plays
a dominant role and, hence, the phenomenological
coefficients for $\psi^3m$ and $\psi m^3$ terms 
have the same smallness as $\gamma$. In such a case, a simple
linear transformation allows to exclude such terms
form the Landau functional without modifying 
significantly the physical meaning of $\psi$ and $m$.

For $\gamma=0$ the functional (\ref{landau}) has a form
commonly found in the theory of phase transitions. Assuming  
that only coefficients $\alpha_{1,2}$ depend on temperature 
and pressure, the energy (\ref{landau}) describes a phase
diagram with two crossing lines of second-order
transitions determined by $\alpha_{1,2}(P,T)=0$.
The transition line from a paramagnetic state
$T_m(P)$ has a kink at the crossing point.
Presence and nature
of extra transitions in the order state, where
$\alpha_1,\alpha_2<0$, depend on quartic terms. 
For $\beta_i <\sqrt{\beta_1\beta_2}$ or
for a weak repulsion between $\psi$ and $m$,
there are two other lines of second-order transitions transition 
emerging from the crossing point. They separate two states
with pure ordering, {\it i.e.},\ 
$\psi\neq 0$, $m=0$ and $\psi=0$, $m\neq 0$, from a mixed
phase with $\psi\neq 0$ and $m\neq 0$.
Thus, the phase diagram in this case has a tetracritical point. 
For $\beta_i>\sqrt{\beta_1\beta_2}$ or
for a strong repulsion between two components,
the mixed phase does not appear. Instead, there is
a single line of first-order transitions in the $P$--$T$ plane
given by $\alpha_1^2\beta_2 = \alpha_2^2\beta_1$, which approaches
the kink (crossing) point from the ordered side, see the left
panel in Fig.~\ref{diagram}.

In the following we discuss effect of nonzero $\gamma$
on the two order parameter functional (\ref{landau}): problem,
which, to our knowledge, has not been considered so far. 
Once $\gamma\neq 0$, the two order parameters appear simultaneously 
on a single transition line given by
$\alpha_1\alpha_2=\gamma^2$.
The transition temperature from a paramagnetic state $T_m(P)$ has 
now a smooth pressure dependence and does not exhibit a kink.
At $P=0$, the induced antiferromagnetic component
behaves as 
\begin{equation}
m\approx -(\gamma/\alpha_2)\psi
\label{MvsPsi}
\end{equation}
for $|\alpha_1|,|\gamma|\ll\alpha_2$. A small coefficient
$\gamma/\alpha_2$ implies weak ordered moments, while $\psi$
gives rise to a large anomaly in the specific heat at $T_m$.
We identify this phase with a small moment antiferromagnetic
(SMAF) phase of URu$_2$Si$_2$.

In order to investigate the possible ordered states and phase
transitions below $T_m(P)$, one has to minimize Eq.~(\ref{landau})
with respect to both $\psi$ and $m$. This gives a system of two
coupled cubic equations, which is easily solved numerically,
but does not allow full analytic solution. Still,
simple analytic arguments can be used to prove 
stability of the first-order transition line for
$\gamma\neq 0$ and $\beta_i>\sqrt{\beta_1\beta_2}$. 
[For $\beta_i<\sqrt{\beta_1\beta_2}$, 
the bilinear term stabilizes the mixed phase 
($\psi\neq 0$ and $m\neq 0$) right below $T_m(P)$.]

Substitution $m\to(\beta_1/\beta_2)^{1/4}\,m$ transforms 
the free-energy to a more symmetric form:
\begin{equation}
F=\alpha_1 \psi^2 + \tilde{\alpha}_2 m^2 + 2\tilde{\gamma}\psi
m+\beta_1(\psi^2 +m^2)^2 + 2\tilde{\beta}_i \psi^2 m^2\,,
\label{fes}
\end{equation}
where $\tilde{\alpha}_2=(\beta_1/\beta_2)^{1/2}\alpha_2$,
$\tilde{\gamma}=(\beta_1/\beta_2)^{1/4}\gamma$, and
$\tilde{\beta}_i=(\beta_1/\beta_2)^{1/2}\beta_i-\beta_1$.
In the new notations the condition for the absence of the mixed phase 
at $\gamma=0$ is $\tilde{\beta}_i>0$, while the position 
of the first-order transitions line in the $P$--$T$ plane is given by
$\alpha_1=\tilde{\alpha}_2$.
Let us cross from the paramagnetic state into the ordered
state along this line taking  $\alpha_1=\tilde{\alpha}_2=\alpha$.
The transformation $\psi=\eta_1-\eta_2$, $m=\eta_1+\eta_2$ 
diagonalizes the quadratic terms in 
Eq.~(\ref{fes}) yielding 
\begin{equation}
F = 2(\alpha+\tilde{\gamma})\eta_1^2 + 
2(\alpha-\tilde{\gamma})\eta_2^2
+ 4\beta_1(\eta_1^2+\eta_2^2)^2 +
2\tilde{\beta}_i(\eta_1^2-\eta_2^2)^2 \ .
\label{fess}
\end{equation}
If we assume, for example, $\gamma<0$, then 
a second order transition takes place at 
$\alpha(T_m)=-\tilde{\gamma}$ from a paramagnetic state into a state with 
$\eta_1^2 = -(\alpha+\tilde{\gamma})/2(2\beta_1+\tilde{\beta}_i)$,
while $\eta_2=0$. For positive $\tilde{\beta}_i$, the last term
in Eq.~(\ref{fess}) disfavors states with $\eta_1^2\neq\eta_2^2$.
Therefore, at sufficiently low temperature there should be another
transition into a state with a nonzero $\eta_2$.
The location of such a
critical point $(T_c,P_c)$ is given by
\begin{equation}
\alpha(T_c) = -2|\tilde{\gamma}|\frac{\beta_1}{\tilde{\beta}_i}= 
 - \frac{2|\gamma|(\beta_1^3\beta_2)^{1/4}}
{\beta_i-\sqrt{\beta_1\beta_2}} \ .
\label{ac2}
\end{equation}  
The ratio of the specific heat jumps at two consecutive
transitions $T_m$ and $T_c$ is
\begin{equation}
\frac{\Delta(C/T)_c}{\Delta(C/T)_m} = \frac{\tilde{\beta}_i}{2\beta_1}\ .
\end{equation}
For $\alpha<\alpha(T_c)$ the two components behave as 
\begin{equation}
\eta_1^2 = -\frac{\tilde{\beta_i}\alpha + 2\tilde{\gamma}\beta_1}
{8\beta_1\tilde{\beta_i}} \ , \ \ 
\eta_2^2 = -\frac{\tilde{\beta_i}\alpha - 2\tilde{\gamma}\beta_1}
{8\beta_1\tilde{\beta_i}} \ . 
\end{equation} 
The relative phase between $\eta_1$ and $\eta_2$ is not fixed,
though solutions $(|\eta_1|,|\eta_2|)$ and $(|\eta_1|,-|\eta_2|)$
describe two essentially different states.
Away from the line $\alpha_1=\tilde{\alpha}_2$, the energy
(\ref{fess}) acquires the extra term 
$2(\alpha_1-\tilde{\alpha}_2)\eta_1\eta_2$, which immediately 
lifts the above two-fold degeneracy and selects either $0$ or 
$\pi$ shift between $\eta_1$ and $\eta_2$ on the two sides 
of $\alpha_1=\tilde{\alpha}_2$.
Consequently, the first order transition line $T_{M}(P)$ is
stable and its position in the $P$--$T$
plane is given by the same equation as for $\gamma=0$.
However, $T_M(P)$ splits from the line of second order phase transitions
$T_m(P)$ and terminates at the critical point determined by Eq.~(\ref{ac2}), 
see the right panel of Fig.~\ref{diagram}.

The two states to the left and to the right from
$T_{M}(P)$ are phases with large $\psi_L=|\eta_1|+|\eta_2|$ 
and small $m_S=|\eta_1|-|\eta_2|$ (SMAF) 
and with small $\psi_S=|\eta_1|-|\eta_2|$ and large 
$m_L=|\eta_1|+|\eta_2|$ (LMAF). 
A relative jump of the ordered antiferromagnetic
moments across $T_M(P)$ is given by
\begin{equation}
\frac{m_L -m_S}{m_L + m_S}= \frac{|\eta_2|}{|\eta_1|} = 
\sqrt{\frac{\alpha-\alpha(T_c)}{\alpha+\alpha(T_c)}} \ .
\end{equation}  
The size of the jump varies continuously along $T_M(P)$ and
vanishes at $P=P_c$. Note, that the distance between
the critical line $T_m(P)$ and the critical point $T_c$ 
given by Eq.~(\ref{ac2}) is proportional to $\gamma$ and may be quite small.
At present, the neutron experiments under hydrostatic 
pressure has failed
to identify the critical end point $(T_c,P_c)$ on the line of
first-order transitions $T_M(P)$.\cite{bourdarot}
We suggest that specific heat measurements can help
to finally resolve the phase diagram of URu$_2$Si$_2$.

\section{Crystal-Field model for induced moments}

The nine-fold degenerate state of U$^{4+}$ ions with the total angular 
momentum $J=4$ is split by a crystalline 
electric field. Following the previous work,\cite{broholm91,sikkema,okuno}
we assume that the ground and the first
excited levels are singlets separated by a crystal field
gap $\Delta$ and that the only nonvanishing matrix
element of the total angular momentum is
$\langle 0| J^z |1\rangle = \mu$. 
Working in the basis of the two lowest levels,
the new pseudo-spin-1/2 operators are defined as
\begin{equation}
S^z |0\rangle = +\frac{1}{2} |0\rangle \ ,\ \ 
S^z |1\rangle = -\frac{1}{2} |1\rangle \ .
\end{equation}
The nonzero component of the angular momentum
operator is expressed in terms of pseudo-spin-1/2 operators
as $J^z = 2\mu S^x$.
Local moments formed by the mixing of two crystal
field levels have, therefore, a very anisotropic nature.
They couple only to a $z$-component of an applied field
and via Ising-like interaction between different sites.
Coupling between local moments and itinerant
carriers is described by an Ising-Kondo interaction \cite{sikkema}
\begin{equation}
\hat{V} = \frac{1}{2}\, I \sum_i J^z_i c^\dagger_{i\alpha} 
\sigma^z_{\alpha\beta} c^{_{}}_{i\beta} \ . 
\label{kondo}
\end{equation}
Below the SDW transition, space modulation of the electron spin
density produces an internal staggered field on uranium sites:
\begin{equation}
H_s({\bf r}_i) = H_s e^{i{\bf Q}{\bf r}_i}\ , \ \ 
H_s = -IM^z_{\bf Q} \sim \psi \ .
\label{e1}
\end{equation}
The estimate of the staggered field from the experimental
data on URu$_2$Si$_2$ is given in section V.
Transformation from Eq.~(\ref{kondo}) to (\ref{e1}) corresponds
to a mean-field approximation, which should be sufficient when
considering two weakly interacting subsystems.

The total crystal-field Hamiltonian in the presence of 
both staggered and uniform fields applied parallel to the 
crystal $\hat{\bf z}$-axis is written in terms of
pseudo-spin operators as 
\begin{eqnarray}
\hat{\cal H} &=& 4\mu^2 \sum_{\langle i,j\rangle} {\cal J}(ij) S^x_i S^x_j
- \Delta \sum_i S^z_i \nonumber \\
&& \mbox{} - 2\mu \sum_i (H  +  
H_se^{i{\bf Q}{\bf r}_i}) S^x_i ,   
\label{HCF}
\end{eqnarray}
where ${\cal J}(ij)$ is a set of exchange constants between local
moments on a body centered tetragonal lattice.

\subsection{Zero-field case}

The crystal field model (\ref{HCF}) in zero applied field has been 
studied by many authors. \cite{bleaney,grover,wang,klenin,jensen} 
The Hamiltonian (\ref{HCF}) without the last term corresponds to a
ubiquitous Ising model in a transverse field. 
In order to make connection with previous works we briefly list 
in this subsection the main results on the crystal-field
model (\ref{HCF}) with $H=H_s=0$.
At zero temperature and in the large gap limit the system
remains in a singlet ground state. The excitation spectrum 
consists of magnetic or Van-Vleck excitons, which are bound states
of the two singlet levels. Energy of Van-Vleck excitons 
is easily found by applying the Holstein-Primakoff representation
to pseudo-spin-1/2 operators. In the harmonic approximation
it is suffice to write
\begin{equation}
S^z_i = \frac{1}{2} - a_i^\dagger a_i \ , \ \ 
S^x_i = \frac{1}{2}\left(a_i^\dagger + a_i\right) \ .
\label{HPT}
\end{equation}
The excitation spectrum is given by
\begin{equation}
\omega_{\bf k} = \sqrt{\Delta(\Delta + 2\mu^2{\cal J}_{\bf k})} \ ,
\label{largeD}
\end{equation}
where ${\cal J}_{\bf k} = \sum_j {\cal J}(ij) e^{i{\bf k}{\bf r}_{ij}}$
is a Fourier transform of the exchange interaction. 
The excitation gap is reduced by magnetic interactions to
\begin{equation}
\Delta_g = \sqrt{\Delta(\Delta-\Delta_c)}\ ,\ \ 
\Delta_c = 2\mu^2 |{\cal J}_{\bf Q}|\ ,
\label{largeDD}
\end{equation}
where the wave-vector $\bf Q$
corresponds to the absolute minimum of ${\cal J}_{\bf k}$. 
Let us emphasize here that neither $\Delta$ nor $\Delta_g$ 
have any relation to
the bulk gap $\Delta_0$, which corresponds to itinerant
magnetic subsystem. 
Quantum fluctuations somewhat renormalize 
the spectrum (\ref{largeD}) at $T=0$ 
and tend to further reduce the critical gap $\Delta_c$.\cite{wang,klenin}
This effect depends, however, on a particular
form of ${\cal J}_{\bf k}$, and for a three-dimensional system does not
exceed 2--3\%. Below, we neglect such quantum corrections. 

If the crystal field splitting $\Delta$ becomes smaller than
$\Delta_c$, the system develops a long-range magnetic order
with a staggered magnetization $\langle J^z_i \rangle = 
2\mu\langle S^x_i\rangle \sim e^{i{\bf Q}{\bf r}_i}$.
In the following
we always assume that magnetic ordering has a 
two-sublattice antiferromagnetic structure, that is 
$e^{2i{\bf Qr}_i}\equiv 1$ or $2{\bf Q}\equiv 0$ as in URu$_2$Si$_2$.
In order to describe a finite-temperature transition into 
ordered state one can use a simple molecular-field 
approximation.\cite{bleaney,grover} 
For this we write 
\begin{equation}
\langle S^x_i\rangle 
= m_s e^{i{\bf Q}{\bf r}_i} \ ,
\end{equation}
where the dimensionless staggered magnetization $m_s$ is determined 
by a self-consistency equation obtained from a single-site
solution:
\begin{equation}
m_s=\frac{2\mu^2|{\cal J}_{\bf Q}|m_s}
{\sqrt{\Delta^2\!+\!16\mu^4{\cal J}^2_{\bf Q}m_s^2}}
\tanh\frac{\sqrt{\Delta^2\!+\!16\mu^4{\cal J}^2_{\bf Q}m_s^2}}{2T} .
\label{mom}
\end{equation}
The transition temperature obtained from the above equation is
\begin{equation}
\frac{\Delta}{T_N} =  \ln \frac{\Delta_c+\Delta}{\Delta_c-\Delta} \ .
\label{TN}
\end{equation}
In the molecular-field approximation the transition temperature 
vanishes as $\Delta\rightarrow\Delta_c-0$
in agreement with Eq.~(\ref{largeDD}).  At zero temperature the
sublattice magnetization is
\begin{equation}
M_{s0} = 2\mu m_{s0} = \mu\: \sqrt{1 - \frac{\Delta^2}{\Delta_c^2}} \ ,
\label{ms0}
\end{equation}
whereas near $T_N$ the antiferromagnetic moments follow
the mean-field temperature dependence:
\begin{equation}
M_s^2 \approx  M_{s0}^2 
\frac{\Delta^2}{T_N\Delta_c - \frac{1}{2}\Delta_c^2+\frac{1}{2}\Delta^2}
\frac{T_{N}-T}{T_N}\ .
\label{msTn}
\end{equation}
The excitation spectrum in the ordered phase at zero temperature 
is found by introducing a staggered canting angle $\varphi$ for 
the two sublattices.\cite{jensen} In the mean-field approximation 
$\cos\varphi=\Delta/\Delta_c$. After transformation to the local
(rotating) frame and application of Eq.~(\ref{HPT}) one finds:
\begin{equation}
\omega_{\bf k} = 2\mu^2 |{\cal J}_{\bf Q}|
\sqrt{1 + \frac{\Delta^2}{\Delta_c^2}\frac{{\cal J}_{\bf k}}{|{\cal J}_{\bf Q}|}} \ .
\label{smallD}
\end{equation}
The above equation shows that
upon approaching the Ising limit $\Delta\ll\Delta_c$, the dispersion
of the longitudinal excitations is gradually diminished.
For more details and discussion see the end of the subsection C. 

Neutron scattering measurements on URu$_2$Si$_2$ yield a moderate value 
of the matrix element of the total angular momentum 
$\mu\simeq 1.2\mu_B$.\cite{broholm91}
A simple explanation of small static moments would be, then, to assume
that $(\Delta_c -\Delta)\ll \Delta$. 
According to Eq.~(\ref{TN}) 
such an assumption also leads to a small 
transition temperature compared to  
the crystal-field level splitting $T_N\ll\Delta$, which is again
in agreement with the experimental observation of
$\Delta\simeq 10$meV.\cite{broholm91}
The above straightforward explanation of small ordered moments
fails, however,
to explain  a large jump of the specific heat at $T_N$.
Indeed, in the molecular-field approximation the specific heat
jump at the transition temperature (\ref{TN}) is
\begin{equation}
\frac{\Delta C}{C} = 2 T_N\: \frac{\Delta_c^2}{\Delta^2}\: 
\left|\frac{dm_s^2}{dT}\right|_{T_N} .
\end{equation}
Using Eq.~(\ref{msTn}) we find in the limit 
$\Delta\rightarrow\Delta_c$:
\begin{equation}
\frac{\Delta C}{C} \approx \frac{\Delta_c^2-\Delta^2}{2\Delta_c^2}
\:\ln\frac{\Delta_c+\Delta}{\Delta_c-\Delta} \ .
\end{equation}
Taking $M_{s0}\approx 0.03\mu_B$, which implies that
$(\Delta_c-\Delta)/\Delta_c\approx 3\times 10^{-4}$, we find
for the specific heat jump $\Delta C/C \approx 2.7\times 10^{-3}$. 
Such a jump is three orders of magnitude
smaller than the experimentally measured jump at the 17.5~K
transition.\cite{maple86}  
Corrections to the molecular-field approximation\cite{wang,klenin}
do not  significantly modify the jump $\Delta C$.
Consequently, it has been concluded that spontaneous ordering
of local moments
on uranium sites cannot explain the phenomenology of
the antiferromagnetic transition in URu$_2$Si$_2$. 
In the next sections we shall consider the model (\ref{HCF})
in the regime of induced local moments, that is 
$\Delta>\Delta_c = 2\mu^2|{\cal J}_{\bf Q}|$ and $H_s\neq 0$.

\subsection{Finite Fields: Mean-field approximation}

The mean-field ansatz for a sublattice magnetization
in the presence of both uniform $H$ and
staggered $H_s$ magnetic fields is given by
\begin{equation}
\langle S^x_i\rangle 
= m_s e^{i{\bf Q}{\bf r}_i} + m_0 \ .
\end{equation}
For a single spin, the mean-field Hamiltonian takes the following form 
\begin{eqnarray}
\hat{\cal H}_{\rm MF} & = & - \Delta S_i^z - S_i^x\bigl[
(2\mu H_s + 4\mu^2|{\cal J}_{\bf Q}|m_s)e^{i{\bf Q}{\bf r}_i} \  
\nonumber \\ 
 & & \left. \mbox{} + 2\mu H - 4\mu^2 {\cal J}_0 m_0\right] , 
\end{eqnarray}
where ${\cal J}_0={\cal J}_{{\bf k}=0}$. 
Calculating  an equilibrium magnetization we find
for two sublattices:
\begin{eqnarray}
&&  m_s \pm m_0  =
\frac{D_\pm}{\sqrt{\Delta^2+4 D_\pm^2}}
\tanh\frac{ \sqrt{\Delta^2+4 D_\pm^2}}{2T} , \nonumber \\
&& D_\pm = \mu(H_s\pm H) + 2\mu^2(|{\cal J}_{\bf Q}|m_s\mp {\cal J}_0m_0) \ .
\label{2self}
\end{eqnarray}
Below, we focus on the case  $\Delta > \Delta_c = 2\mu^2|{\cal J}_{\bf Q}|$,
when there is no magnetic ordering in the subsystem of local
moments down to $T=0$
in the absence of both external and internal
fields. 
For weak staggered field, linearization of  
Eq.~(\ref{2self}) in $H_s$ and $m_s$ at $H=0$ yields
\begin{equation}
m_s = \frac{\mu H_s \tanh(\Delta/2T)}{\Delta-\Delta_c\tanh(\Delta/2T)} \ .
\end{equation}
In accordance with the phenomenological formula (\ref{MvsPsi}) of 
section II, weak local moments are proportional to 
the primary (SDW) order parameter.
At zero temperature the dimensional staggered moments are
\begin{equation}
M_{s0} = 2\mu \frac{\mu H_s}{\Delta-\Delta_c} \ .
\label{Ms0}
\end{equation}
The above equation allows to estimate the staggered field
in URu$_2$Si$_2$ from the available experimental data, 
see section V.

The effect of a uniform field on induced local moments 
is considered, for simplicity, for $T=0$ only. 
In this case expansion of Eq.~(\ref{2self})
to linear order
in $m_s$ and $H_s$ yields 
\begin{equation}
m_s = \frac{\mu H_s}{\Delta(1+ 4\mu^2\widetilde{H}^2/\Delta^2)^{3/2} -
\Delta_c} \ ,
\end{equation}
where an effective field $\widetilde{H} = H - 2\mu {\cal J}_0 m_0$
is determined self-consistently from 
\begin{equation}
\widetilde{H} = H - \frac{2\mu^2{\cal J}_0 \widetilde{H}}{(\Delta^2 + 
4\mu^2 \widetilde{H}^2)^{1/2}} \ .
\label{selfH}
\end{equation}
According to arguments of section III, the SDW contribution
to the magnetic Bragg peaks is negligible due to a small
form-factor. Then, the measured intensity of Bragg reflections 
is proportional to $m_s^2$. Suppression of an SDW 
order parameter with an external field can be described by a simple formula
$\psi^2\propto(1-H^2/H_c^2)$ or $H_s^2=H_{s0}^2(1-H^2/H_c^2)$, 
where $H_c\simeq 40$~T is 
a metamagnetic field in URu$_2$Si$_2$.\cite{harrison}
The magnetic Bragg peak intensity is
\begin{equation}
I_{\bf Q} \sim m_s^2 = \frac{\mu^2 H_{s0}^2(1- H^2/H_c^2)}
{[\Delta(1+4\mu^2\widetilde{H}^2/\Delta^2)^{3/2}-\Delta_c]^2} \ .
\label{ms2}
\end{equation}
This zero-temperature result should be compared to the analogous 
formulas valid near $T_m$, which have been derived 
in the previous works \cite{shah,bourdarot03}
from the Landau free-energy functional.
Though different in the details, the two limits exhibit an inflection point
in $I_{\bf Q}(H)$.
The staggered magnetization remains finite until 
$H_c$, when the primary order parameter is suppressed to zero.
The ordered moments are, however, substantially reduced 
compared to its zero-field value at significantly smaller
magnetic field. Indeed, expanding Eqs.~(\ref{selfH}) and (\ref{ms2})
to the first order in $H^2$ we obtain
\begin{equation}
\frac{m_s^2(H)}{m_s^2(0)} \approx  1 - H^2 \left(\frac{1}{H_c^2} +
\frac{12\mu^2\Delta}
{(\Delta\!-\!\Delta_c)(\Delta\!+\!2\mu^2{\cal J}_0)^2}\right).
\end{equation}

For the completeness we also note that the ferromagnetic
component of the induced magnetic moments is given by
\begin{equation}
m_0=\frac{\mu H}{\Delta+2\mu^2 {\cal J}_0}
\label{m0}
\end{equation}
for fields smaller than $H^*=(\Delta+2\mu^{2}{\cal J}_0)/2\mu$.
Above this field the ferromagnetic component remains constant until
a metamagnetic transition related to a crossing with higher
energy crystal-field levels.  
The uniform component of the induced moments $m_0$ 
should be measurable from a 
magnetic contribution to the nuclear Bragg peaks.

\subsection{Finite Fields: Energy Spectrum}

We start with the case $H\neq 0$, $H_s=0$, since an
effective $H_s$ in URu$_2$Si$_2$ should be quite small. Partial
polarization of magnetic moments (pseudo-spins) along $\hat{z}$
($\hat{x}$) axis is taken into account by rotation of
pseudo-spins from a laboratory frame to a local (primed) frame:
\begin{eqnarray}
S_i^x & = & S_i^{x'} \cos \varphi + S_i^{z'} \sin\varphi \ , \nonumber \\
S_i^z & = & -S_i^{x'} \sin \varphi + S_i^{z'} \cos\varphi \ .
\label{local1}
\end{eqnarray}
In the transformed frame and omitting primes
the Hamiltonian (\ref{HCF}) takes the 
following form:
\begin{eqnarray}
\hat{\cal H} & = & 4\mu^2 \sum_{\langle i,j\rangle} {\cal J}(ij) \Bigl[
S^x_i S^x_j \cos^2\!\varphi + S^z_i S^z_j \sin^2\!\varphi
\nonumber \\
& & \mbox +
(S^x_i S^z_j + S^z_i S^x_j) \sin\varphi\cos\varphi \Bigr] \nonumber \\
&&\mbox{}-\sum_i\Bigl[\bigl(\Delta\cos\varphi+2\mu H\sin\varphi\bigr)S^z_i
\label{Hunif} \\
& & \ \ \ \ \  \ \ \ \ \mbox{} +
\bigl(2\mu H\cos\varphi-\Delta\sin\varphi\bigr) S^x_i \Bigr] .
\nonumber
\end{eqnarray}
The boson representation (\ref{HPT}) of the pseudo-spin operators
is applied to the above Hamiltonian  and the
rotation angle $\varphi$ is determined from the condition of vanishing
linear terms in $a_i$ and $a_i^\dagger$:
\begin{equation}
\Delta\tan\varphi + 2\mu^2 {\cal J}_0 \sin\varphi = 2 \mu H \ .
\label{angle}
\end{equation}
At small fields $\varphi \approx 2\mu H/(\Delta+2\mu^2{\cal J}_0)$.

The harmonic part of the Hamiltonian (\ref{Hunif}) after the Fourier
transformation becomes
\begin{eqnarray}
&& \hat{\cal H}_2  =  \sum_{\bf k} a_{\bf k}^\dagger a_{\bf k}
[\Delta\cos\varphi +2 \mu H\sin\varphi - 2\mu^2\sin^2\!\varphi {\cal J}_0
\nonumber  \\
 & &  + \mu^2\cos^2\!\varphi {\cal J}_{\bf k}] 
  + \frac{1}{2}\:\mu^2\cos^2\!\varphi {\cal J}_{\bf k}
( a_{\bf k}a_{-{\bf k}}\! +\! 
 a_{\bf k}^\dagger a_{-{\bf k}}^\dagger ). 
\end{eqnarray}
The $\bf k$-independent term is simplified with the help 
of Eq.~(\ref{angle}) to $\Delta/\cos\varphi$ and after applying 
the Bogoliubov transformation we obtain the following
field dependence of the exciton spectrum:
\begin{equation}
\omega_{\bf k}^2 = \frac{\Delta^2}{\cos^2\varphi} +
2\mu^2 {\cal J}_{\bf k}\Delta\cos\varphi \ .
\label{spectrumH}
\end{equation}
The gap at ${\bf k}={\bf Q}$ increases quadratically with
magnetic field:
\begin{equation}
\Delta_g^2(H) \approx \Delta(\Delta-\Delta_c) + 
\frac{2\mu^2 H^2}{(\Delta+2\mu^2 {\cal J}_0)^2} \: 
\Delta (2\Delta+\Delta_c)\ .
\label{GapH2}
\end{equation}
The parabolic law for the field dependence of the gap 
has recently been measured in neutron scattering experiments. 
\cite{bourdarot03} For an arbitrary wave-vector the field
dependence of the exciton energy is 
\begin{equation}
\omega_{\bf k}^2(H) \approx \omega_{\bf k}^2(0) +
\frac{4\mu^2 H^2\Delta}{(\Delta+2\mu^2 {\cal J}_0)^2} \:\left[\Delta - 
\mu^2 {\cal J}_{\bf k} \right] .
\end{equation}
The field dependence changes its sign, {\it i.e.\/}, 
the energy starts to decrease
with magnetic field, for the wave-vectors such that 
$\mu^2 {\cal J}_{\bf k}>\Delta$. In terms of zero field frequencies
this is equivalent to $\omega_{\bf k} > \sqrt{3}\Delta$.
In the region in the Brillouin zone where $\omega_{\bf k}\approx 
\sqrt{3}\Delta$ the field dependence of the spectrum becomes vanishingly
small. Experimentally, a drastic change in the field dependence
has been observed between  ${\bf k}={\bf Q}=(1,0,0)$ and
${\bf k} = (1.4,0,0)$. The present theory explains
a qualitative difference in the field
response of the two types of excitons. Further inelastic neutron measurements
on URu$_2$Si$_2$ should allow a detailed comparison with our theory and 
extraction of microscopic parameters from experimental data. 

If both staggered and uniform fields are present, the derivation
of the spectrum becomes a bit more complicated.
One has to explicitly introduce two types of bosons
$a_i$ and $b_i$ for two antiferromagnetic sublattices
and to calculate spectrum in the magnetic Brillouin zone, which is
twice smaller than an original lattice Brillouin zone used
above. The transformation to local (primed) axes from the laboratory
frame is given by
\begin{eqnarray}
S_i^x & = & S_i^{x'}\cos\varphi_i + S_i^{z'}
e^{i{\bf Q}{\bf r}_i} \sin\varphi_i \ , \nonumber \\
S_i^z & = & -S_i^{x'} e^{i{\bf Q}{\bf r}_i}\sin \varphi_i + 
S_i^{z'} \cos\varphi_i \ ,
\label{local2}
\end{eqnarray}
where two angles $\varphi_i=\varphi_1$, for $e^{i{\bf Q}{\bf r}_i}=1$ and
$\varphi_i=\varphi_2$, for $e^{i{\bf Q}{\bf r}_i}=-1$ describe different
response of the two sublattices.  
The angles are determined by
\begin{eqnarray*}
 & \Delta\tan\varphi_1 +  2\mu^2 {\cal J}_1 \sin\varphi_1 
= 2 \mu (H_s\!+\!H) +   2\mu^2 {\cal J}_2 \sin\varphi_2 , & \\
 &\Delta\tan\varphi_2 +  2\mu^2 {\cal J}_1 \sin\varphi_2 
= 2 \mu (H_s\!-\!H) +   2\mu^2 {\cal J}_2 \sin\varphi_1 . &
\end{eqnarray*}
here we defined separate summation of exchange
constants over the same 
${\cal J}_1 = \sum_{i,j\in A} {\cal J}(ij)$ 
and the different sublattices
${\cal J}_2 = \sum_{i\in A,j\in B} {\cal J}(ij)$. 

Performing the same steps as in the case of $H_s=0$ we in the end find
\begin{eqnarray}
\omega_{\bf k}^{\pm 2} & = & 
\frac{1}{2}(\omega^2_{1{\bf k}} + \omega^2_{2{\bf k}}) \nonumber \\
&& \pm \sqrt{ \frac{1}{4}(\omega^2_{1{\bf k}} - \omega^2_{2{\bf k}})^2
+ 4\mu^4 {\cal J}_{2{\bf k}}^2\Delta^2\cos\varphi_1\cos\varphi_2} \ , \nonumber \\
& & \omega^2_{1,2{\bf k}} = \frac{\Delta^2}{\cos^2\varphi_{1,2}} + 
2\mu^2 {\cal J}_{1{\bf k}}\Delta \cos\varphi_{1,2} \ .
\end{eqnarray}
Here, the Fourier transforms are given by
${\cal J}_{1{\bf k}} = \sum_{i,j\in A} {\cal J}(ij) e^{i{\bf k}{\bf r}_{ij}}$ 
and 
${\cal J}_{2{\bf k}} = \sum_{i\in A,j\in B} {\cal J}(ij)e^{i{\bf k}{\bf r}_{ij}}$. 
The characteristic feature of this spectrum is a small jump
between two branches of excitations 
$\omega_{\bf k}^+$  and $\omega_{\bf k}^-$  
at the magnetic Brillouin zone boundary, where ${\cal J}_{2{\bf k}}\equiv 0$. 
In URu$_2$Si$_2$ ($H_s\neq 0$) such a jump is 
induced by external magnetic field $H\sim H_s$ and becomes
negligible again for $H\gg H_s$, when the above expression
Eq.~(\ref{spectrumH}) is valid.

Finally, let us comment on the longitudinal polarization
of magnetic excitons detected experimentally.\cite{broholm91}
In the harmonic approximation, 
the dynamic structure factor $S^{zz}({\bf r},\tau) =
\langle J^z_i(t) J^z_{i+{\bf r}}(t+\tau)\rangle$
is expressed via 
pseudo-spin operators as
\begin{equation}
S^{zz}({\bf r},\tau) \approx 4\mu^2 \cos^2\varphi 
\langle S^x_i(t) S^x_{i+{\bf r}}(t+\tau)\rangle \ .
\label{Szz}
\end{equation}
In a weakly polarized Van-Vleck paramagnet for $\Delta\sim\Delta_c$ 
one has $\cos\varphi_{1,2} \approx 1$.
Therefore, transverse `spin-wave-type modes' in the pseudo-spin 
representation correspond to longitudinal polarization
of magnetic excitons.
The omitted terms in Eq.~(\ref{Szz}), such as 
$\langle S^z_i S^z_{i+{\bf r}}\rangle$ and
$\langle S^x_i S^z_{i+{\bf r}}\rangle$, describe 
a higher energy two-magnon continuum and its interaction
with single-particle states. These terms do not modify
the conclusion about longitudinal polarization of
single-particle excitations.
In the opposite limit $\Delta\ll\Delta_c$ in a phase with large 
antiferromagnetic moments $\cos\varphi_{1,2}\rightarrow 0$,
and the longitudinal dynamic structure factor
$S^{zz}({\bf q},\omega)$ does not have contribution
from magnetic excitons.
This explains why the neutron scattering measurements\cite{bourdarot03} 
failed
to observe the magnetic excitations above the first-order
transition at $P_M=5$~kbar.

\begin{figure}[t]
\begin{center}
\includegraphics[width=0.95\columnwidth]{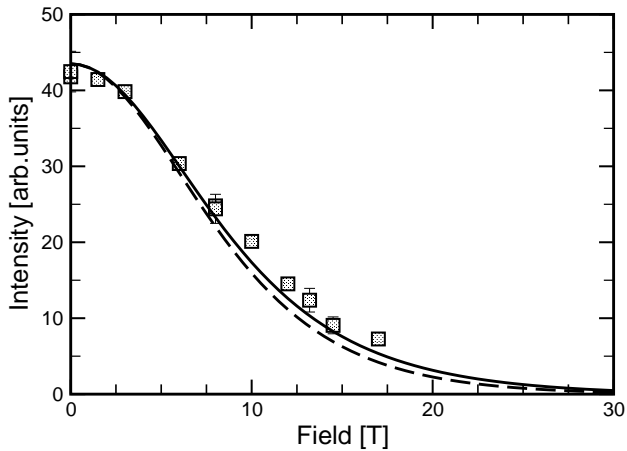}
\end{center}
\caption{
Field dependence of the intensity of the 
antiferromagnetic Bragg peak at ${\bf Q}=(1,0,0)$.
Points are the experimental data.\cite{bourdarot03}
Lines are theoretical curves described by Eq.~(\ref{ms2}) with
$\Delta=6$~meV (full line) and 
$\Delta=3$~meV (dashed line).
}
\label{Bragg}
\end{figure}

\section{Comparison with experiment}

Theoretical predictions of the above section can be directly
compared with the available experimental data.
Specifically, let us consider the field dependence of the
intensity of the magnetic Bragg peak. 
The metamagnetic transition in URu$_2$Si$_2$ at
$H_c \simeq 40$~T can be chosen as the critical field for 
the spin-density wave.
The field dependence of the Bragg peak intensity
described by Eqs.~(\ref{selfH}) and (\ref{ms2}) is, then, determined
by three microscopic parameters: $\Delta$, $\Delta_c=2\mu^2|{\cal J}_{\bf Q}|$,
and $2\mu^2{\cal J}_0$. 
The last two parameters are fixed by using the experimental data 
\cite{bourdarot03} for the excitation gap
$\Delta_g=\sqrt{\Delta(\Delta-\Delta_c)}\approx 1.59$~meV and
its dependence on applied magnetic field.
In this way we are left with only one free parameter: 
the crystal-field splitting $\Delta$. 

The two theoretical curves for $\Delta=3$~meV 
($\Delta_c=2.2$~meV and $2\mu^2{\cal J}_0=1.3$meV)
and $\Delta=6$~meV
($\Delta_c=5.57$~meV and $2\mu^2{\cal J}_0=2.9$meV)
are presented in Fig.~\ref{Bragg} together with the experimental
data.\cite{bourdarot03} 
Both curves exhibit behavior with an inflection point. 
The larger value of the gap gives better agreement with the
experimental results. 
For $\Delta=6$~meV the top
of the exciton band at $H=0$ calculated 
from Eq.~(\ref{largeD}) corresponds 
to $\omega_0\approx 7.3$~meV. 
The internal staggered field estimated from Eq.~(\ref{Ms0})
in this case is $H_s \approx 0.08$~T, which is indeed much smaller than 
applied magnetic fields and justifies the used approximations.

If we further increase
$\Delta$, the theoretical dependence for $I_{\bf Q}(H)$
with the above two constraints practically saturates
at the position given by $\Delta=6$~meV curve.
Thus, while we can definitely
exclude smaller values $\Delta<6$~meV for the crystal-field
level splitting, the larger values $\Delta>6$~meV  are equally
possible. For example, for $\Delta=10$~meV, which has been 
suggested on the basis of the early neutron scattering measurements,
\cite{broholm91} the parameters obtained from the fits are
$\Delta_c=9.7$~meV, $2\mu^2{\cal J}_0=4.9$meV, and 
$\omega_0\approx 12.2$~meV.
At present, there is no agreement on the value of $\omega_0$
between the two groups of inelastic neutron measurements.
\cite{broholm91,bourdarot03,bourdarotPhD}
Additional more precise neutron scattering investigation of URu$_2$Si$_2$
should greatly help to settle this dispute. 

\section{Conclusions}

We have presented a theoretical model to describe unusual magnetism
in URu$_2$Si$_2$ below $T_m=17.5$~K, which combines tiny ordered
moments $\mu \sim 0.03\mu_B$ with a large specific heat anomaly
at the transition point. At ambient pressure, the transition is driven 
by an SDW instability in the itinerant subsystem, which also induces
weak ordering of local moments on uranium sites. We argue that such 
a low-$T_c$ spin-density wave has a small form-factor and does
not contribute significantly to the neutron scattering,
which essentially probe the localized subsystem.
Phenomenologically, the phase diagram of URu$_2$Si$_2$ is described
by the two order parameter functional (\ref{landau}), which is consistent
with the first-order transition into a state with large antiferromagnetic
moments. The microscopic origin of a strong repulsion between two
order parameters of the same symmetry needs further clarification.
In our view such a behavior may result from a strong renormalization
of the RKKY type interaction between the local moments by 
a rather large SDW gap, which opens over a half of the Fermi surface.

In our discussion we have assumed following the previous works
\cite{broholm91,sikkema,okuno} that uranium ions are
in the $^3H_4$ ground state. The validity of such an assumption
needs further clarification. Also, the high field behavior with
an itinerant electron metamagnetic transition \cite{harrison}
would be an interesting
test for the dual model and its extension suggested in in the present
work. 
Another open question is temperature evolution
of the crystal-field excitations. We believe that the experimentally
observed disappearance of the magnetic excitons above the transition
temperature \cite{broholm91,bourdarot03} is largely related 
to the closeness of two energy scales:
$T_m=17.5$~K and $\Delta_g = 1.6$~meV.

The analysis presented in section IVa may be also relevant to UPt$_3$,
another heavy-fermion compound with tiny antiferromagnetic moments,
for review see Ref.~\onlinecite{vandijk02}.
This material does not have apparent anomalies in thermodynamic
and kinetic properties at $T_m\approx 5$~K, though the neutron
diffraction experiments have detected 
small antiferromagnetic moments $\mu\sim 0.02\mu_B$.
In a possible scenario for UPt$_3$, there is no a SDW instability 
in the conduction subsystem.
The phase transition is driven by the RKKY or the superexchange
interaction between local moments, which only slightly exceeds the
critical value for zero-temperature antiferromagnetic ordering 
determined by a crystal-field level splitting.
While the crystal level structure is not precisely known
for UPt$_3$, the estimates for the specific heat anomaly
given in the end of section IVa
should be generally valid.
Thus, the small ordered antiferromagnetic moments can be reconciled
with the absence of large anomalies 
at the transition point.
The pressure effect on antiferromagnetic ordering in UPt$_3$
also agrees with Eqs.~(\ref{TN}) and (\ref{ms0}), which predict
a much faster square-root suppression of zero-temperature moments
compared to a slow logarithmic decrease of the transition temperature.

\section{Acknowledgements}

It is pleasure to express our gratitude to F. Bourdarot and B. F\aa k
for the numerous stimulating discussions.
We would also like to thank to M. R. Norman and T. M. Rice for valuable
comments.


\begin{thebibliography}{99}

\bibitem{palstra85} 
T. T. M. Palstra, A. A. Menovsky, J. van den Berg, A. J. Dirkmaat,
P. H. Kes, G. J. Nieuwenhuys, and J. A. Mydosh,
Phys. Rev. Lett.  {\bf 55}, 2727 (1985).  

\bibitem{schlabitz} 
W. Schlabitz, J. Baumann, B. Pollit, U. Rauchschwalbe, H. M. Meyer, 
U. Ahlheim, and C. D. Bredl, Z. Phys. B {\bf 62}, 171 (1986).  

\bibitem{maple86} 
M. B. Maple, J. W. Chen, Y. Dalichaouch, T. Kohara, C. Rossel, 
M. S. Torikachvili, M. W. McElfresh, and J. D. Thompson, 
Phys. Rev. Lett.  {\bf 56}, 185 (1986). 

\bibitem{palstra86} 
T. T. M. Palstra, A. A. Menovsky, and J. A. Mydosh, 
Phys. Rev. B {\bf 33}, R6527 (1986).  

\bibitem{visser}
A. de Visser, F. E. Kayzel, A. A. Menovsky, J. J. M. France, 
J. van den Berg, and G. J. Nieuwenhuys,
Phys. Rev. B {\bf 34}, R8168 (1986).

\bibitem{broholm87}
C. Broholm, J. K. Kjems, W. J. L. Buyers, P. Matthews, T. T. M. Palstra, 
A. A. Menovsky, and J. A. Mydosh,
Phys. Rev. Lett. {\bf 58}, 1467 (1987).  

\bibitem{mason90} 
T. E. Mason, B. D. Gaulin, J. D. Garrett, Z. Tun,
W. J. L. Buyers, and E. D. Isaacs, Phys. Rev. Lett. {\bf 65}, 3189 (1990). 

\bibitem{broholm91}
C. Broholm, H. Lin,  P. T. Matthews, T. E. Mason,  W. J. L. Buyers,
M. F. Collins,  A. A. Menovsky, J. A. Mydosh, and J. K. Kjems, 
Phys. Rev.B {\bf 43}, 12809 (1991).

\bibitem{fak96}
B. F\aa k, C. Vettier, J. Flouquet, F. Bourdarot, S. Raymond, A. Verni\`ere,
P. Lejay, Ph. Boutrouille, N. R. Bernhoeft, S. T. Bramwell, R. A. Fisher,
N. E. Phillips,
J. Mag. Mag. Mater. {\bf 154}, 339 (1996).

\bibitem{isaacs} 
E. D. Isaacs, D. B. McWhan, R. N. Kleiman, D. J. Bishop, G. E. Ice, 
P. Zschack, B. D. Gaulin, T. E. Mason, J. D. Garrett, and 
W. J. L. Buyers, Phys. Rev. Lett. {\bf 65}, 3185 (1990). 

\bibitem{nieuwenhuys}  
G. J. Nieuwenhuys, Phys. Rev. B {\bf 35}, 5260 (1987).

\bibitem{gorkov}
L. P. Gor'kov, Europhys. Lett. {\bf 16}, 303 (1991);
L. P. Gor'kov and A. Sokol,  Phys. Rev. Lett. {\bf 69}, 2586 (1992).

\bibitem{ramirez}
A. P. Ramirez, P. Coleman, P. Chandra, E. Br\"uck, A. A. Menovsky, Z. Fisk, 
and E. Bucher,  
Phys. Rev. Lett.  {\bf 68}, 2680 (1992).

\bibitem{barzykin}
V. Barzykin and L. P. Gor'kov, 
Phys. Rev. Lett. {\bf 70}, 2479 (1993).

\bibitem{agterberg} 
D. F. Agterberg and M. B. Walker,  Phys. Rev. B {\bf 50}, 563 (1994).

\bibitem{santini} 
P. Santini and G. Amoretti, Phys. Rev. Lett.  {\bf 73}, 1027 (1994);
M. B. Walker and W. J. L. Buyers,  {\it ibid\/}.\  {\bf 74}, 4097 (1995);
P. Santini and G. Amoretti, {\it ibid\/}.\ {\bf 74}, 4098 (1995).

\bibitem{sikkema}
A. E. Sikkema, W. J. L. Buyers, I. Affleck, and J. Gan, 
Phys. Rev. B {\bf 54}, 9322 (1996).

\bibitem{okuno}
Y. Okuno and K. Miyake, J. Phys.  Soc.  Jpn.  {\bf 67}, 2469 (1998).

\bibitem{ikeda} 
H. Ikeda and Y. Ohashi, Phys. Rev. Lett. {\bf 81}, 3723 (1998).

\bibitem{shah}
N. Shah, P. Chandra, P. Coleman, and J. A. Mydosh,  
Phys. Rev. B {\bf 61}, 564 (2000).


\bibitem{chandra}
P. Chandra, P. Coleman, J. A. Mydosh, and V. Tripathi, 
Nature {\bf 417}, 831 (2002); 
P. Chandra, P. Coleman and J. A. Mydosh, 
Physica B {\bf 312-313}, 397 (2002).

\bibitem{fazekas}
A. Kiss and P. Fazekas, {\tt e-print:\ cond-mat/0411029}.

\bibitem{fomin}
I. A. Fomin, unpublished (2004).

\bibitem{bourdarot03}
F. Bourdarot, B. F\aa k, K. Habicht, and K. Prokes,  
Phys. Rev. Lett. {\bf 90}, 067203 (2003).

\bibitem{amitsuka}
H. Amitsuka, M. Sato, N. Metoki, M. Yokoyama, K. Kuwahara,
T. Sakakibara, H. Morimoto, S. Kawarazaki, Y. Miyako, and J. A. Mydosh,
Phys. Rev. Lett. {\bf 83}, 5114 (1999).

\bibitem{motoyama}
G. Motoyama, T. Nishioka, and N. K. Sato,
Phys. Rev. Lett.  {\bf 90}, 166402 (2003).

\bibitem{bourdarot}
F. Bourdarot, B. F\aa k, V. P. Mineev, M. E. Zhitomirsky, N. Kernavanois, 
S. Raymond, P. Burlet, F. Lapierre, P. Lejay, J. Flouquet, 
{\tt e-print:\ cond-mat/0312206\/}, and to be published in Physica B.

\bibitem{matsuda01}
K. Matsuda, Y. Kohori, T. Kohara, K. Kuwahara, and H. Amitsuka,
Phys. Rev. Lett.  {\bf 87}, 087203 (2001).

\bibitem{matsuda03}
K. Matsuda, Y. Kohori, T. Kohara, K. Kuwahara, and T. Matsumoto, 
J. Phys.: Condens.  Matter {\bf 15}, 2363 (2003).

\bibitem{norman}
M. R. Norman, T. Oguchi, and A. J. Freeman, 
Phys. Rev. B {\bf 38}, 11\,193 (1988).

\bibitem{mzh} 
M. E. Zhitomirsky and V.-H. Dao, Phys. Rev. B {\bf 69}, 054508 (2004).

\bibitem{fisher90}
R. A. Fisher, S. Kim, Y. Wu, N. E. Phillips, M. W. McElfresh, 
M. S. Torikachvili, and M. B. Maple,
Physica B {\bf 163}, 419 (1990).

\bibitem{ohkuni}
H. Ohkuni, Y. Inada, Y. Tokiwa, K. Sakurai, R. Settai, T. Honma, Y. Haga, 
E. Yamamoto, Y. Onuki, H. Yamagami, S. Takahashi, and T. Yanagisawa, 
Phil. Mag. B {\bf 79}, 1045 (1999).

\bibitem{mentink}
S. A. M. Mentink, T. E. Mason, S. S\"ullow, G. J. Nieuwenhuys, 
A. A. Menovsky, J. A. Mydosh, and J. A. A. J. Perenboom,
Phys. Rev. B {\bf 53}, R6014 (1996).

\bibitem{vandijk}
N. H. van Dijk, F. Bourdarot, J. C. P. Klaasse, I. H. Hagmusa, 
E. Br\"uck, and A. A. Menovsky, 
Phys. Rev. B {\bf 56}, 14493 (1997).

\bibitem{jaime}
M. Jaime, K. H. Kim, G. Jorge, S. McCall, and J. A. Mydosh,
Phys. Rev. Lett. {\bf 89}, 287201 (2002).

\bibitem{mckenzie} 
R. H. McKenzie, {\tt e-print:\ cond-mat/9706235}.

\bibitem{gruner} G. Gr\"uner, {\it Density Waves in Solids} 
(Perseus Publishing,
Cambridge Massachusetts, 1994).

\bibitem{skalski}
S. Skalski, O. Betbeder-Matibet, and P. R. Weiss,
Phys. Rev. {\bf 136}, A1500 (1964).

\bibitem{varma}
C. M. Varma, Phys. Rev. Lett. {\bf 55}, 2723 (1985).

\bibitem{norman87}
M. R. Norman, Phys. Rev. Lett. {\bf 59}, 232 (1987).

\bibitem{bleaney}
B. Bleaney, Proc. Roy. Soc. (London) {\bf 276A}, 19 {1963}.

\bibitem{grover}
B. Grover, Phys. Rev. {\bf 140}, A1944 (1965).

\bibitem{wang}
Y.-L. Yang and B. R. Cooper, Phys. Rev. {\bf 172}, 539 (1968);
{\it ibid\/}. {\bf 185}, 696 (1969).

\bibitem{klenin}
M. A. Klenin and J. A. Hertz, Phys. Rev. B {\bf 14}, 3024 (1976).

\bibitem{jensen}
J. Jensen and A. R. Mackintosh, {\it Rare Earth Magnetism: Structure and
Excitations\/}, (Clarendon Press, Oxford, 1991).

\bibitem{harrison}
N. Harrison, M. Jaime, and  J. A. Mydosh,
Phys. Rev. Lett.  {\bf 90}, 096402 (2003).

\bibitem{bourdarotPhD}
F. Bourdarot, Ph. D. Thesis, unpublished (1994).

\bibitem{vandijk02}
N. H. van Dijk, P. Rodi\`ere, B. F\aa k, A. Huxley, and J. Flouquet,  
Physica B {\bf 319}, 220 (2002).



\end{thebibliography}
\end{document}